\documentclass[11pt,a4paper]{article}

\usepackage[utf8]{inputenc}
\usepackage[T1]{fontenc}
\usepackage{newtxtext,newtxmath}
\usepackage[textwidth=14cm,vmargin=1in]{geometry}
\usepackage{microtype}
\usepackage{xcolor}
\usepackage{enumitem}
\usepackage[autostyle]{csquotes}
\usepackage[backend=biber,style=numeric-comp,date=year]{biblatex}
\usepackage{authblk}
\usepackage[hidelinks]{hyperref}

\newcommand{\header}[1]{\vspace{2mm}\noindent\textbf{#1.}}
\newcommand{\headerl}[1]{\vspace{2mm}\noindent\textit{#1.}}

\title{ML Engineering Agents -- \\A Data Management Perspective}
\author{Arnab Phani}
\author{Elias Strauss}
\author{Tim Schuster}
\author{Sebastian Schelter}
\affil{BIFOLD \& TU Berlin, Germany\\
  \texttt{\{arnab.phani, elias.strauss, tim.schuster\}@tu-berlin.de}, \texttt{schelter@tu-berlin.de}}
\date{}

\begin{document}
\maketitle
\vspace{-2.5em}
\begin{abstract}
\noindent Machine learning engineering (MLE) agents are emerging as a new paradigm for automating end-to-end ML pipeline development. Using large language models for code generation and predictive performance as a machine-gradable feedback signal, these agents iteratively synthesize, execute, and refine pipeline code. Yet current systems are largely evaluated on pre-built, single-table benchmarks that abstract away key data management challenges.
This tutorial introduces the foundations of MLE agents and highlights research opportunities for the database community. We cover how agents combine program synthesis with iterative and evolutionary optimization, and discuss open challenges in efficient pipeline execution, data discovery and augmentation over heterogeneous sources, reliability and debugging, as well as fairness and compliance. We connect these challenges to existing data management techniques and conclude with a hands-on session using contemporary MLE agents.
\end{abstract}
\vspace{1em}

\section{Topic of the Proposed Tutorial}

Machine learning (ML) on tabular data is central to data science research and practice~\cite{van2024tabular}. In real-world applications, data comes from heterogeneous sources such as data lakes, warehouses, logs, and external APIs, and must be joined, cleaned, augmented~\cite{chepurko2020arda,abedjan2026data,hollmann2023large}, and encoded before it can be used by predictive models.

\header{Machine learning pipelines} This preparation is implemented through end-to-end \emph{ML pipelines}~\cite{polyzotis2017data}. Such pipelines span data integration, prediction, debugging, and postprocessing~\cite{polyzotis2019data,karlavs2022data,cappuzzo2025retrieve}, combine code from many libraries~\cite{psallidas2022data}, and can be expensive to execute in production~\cite{xin2021production}. Developing them is therefore tedious and error-prone, requiring substantial engineering and domain expertise as well as repeated revisions to improve predictive performance. Accordingly, recent industry surveys identify data quality and pipeline consistency as major obstacles to deploying ML-based applications~\cite{anaconda2025stateofdatascience}.

\header{The rise of machine learning engineering agents} Large language models (LLMs) are reshaping data-centric software development through advances in code synthesis and program understanding. A key driver is their ability to optimize programs for \textit{machine-gradable problems}~\cite{yang2024large}, where a cheap, well-defined feedback metric can guide iterative improvement. This paradigm has for example enabled advances in scientific and algorithmic discovery (exemplified by Google's AlphaEvolve~\cite{novikov2025alphaevolve}), workload-specific database engine synthesis~\cite{wehrstein2026bespoke,lao2026demonstrating}, and kernel optimization for emerging AI accelerators~\cite{zhang2026accelopt}.

\noindent ML pipelines are a natural fit for this paradigm because their predictive performance provides a direct optimization objective, making pipeline generation machine-gradable. This has led to the emergence of \textit{machine learning engineering (MLE) agents}, which aim to automate the generation of tabular prediction pipelines from natural-language task descriptions. These agents use LLMs in iterative loops to generate and refine pipeline code, treating pipeline construction as a program optimization problem. They typically combine LLM-based code generation with evolutionary search methods, such as Monte-Carlo tree search~\cite{browne2012survey} or MAP-Elites~\cite{mouret2015illuminating}, using validation performance as the fitness signal. Representative systems include AIDE~\cite{jiang2025aide}, Amazon's MLZero~\cite{fang2026mlzero}, Google's MLE-Star~\cite{nam2026mle}, and MLEvolve~\cite{du2026mlevolve}. Several conceptually extend earlier work from the data management community~\cite{fathollahzadeh2025catdb,shang2019democratizing,lopez2023alphad3m,li2023diffprep,hilprecht2023diffml,siddiqi2023saga}. However, current agents are primarily evaluated on simplified benchmarks such as MLE-bench, which is based on pre-built datasets derived from Kaggle competitions~\cite{chan2025mle,toledo2026ai} and thereby ignores key data management challenges.

\header{Data management challenges for MLE agents} (Semi-)autonomous agents are emerging as a new ``class of users'' of data systems. In response, the data management community is beginning to redesign systems to be ``agent-first''~\cite{liu2025supporting,luo2026data}, with work spanning agentic text-to-SQL~\cite{hong2025next}, semantic query processing~\cite{zhu2026beyond}, data science~\cite{kddcup2026dataagents}, and deep research~\cite{russo2025deep}. We argue that MLE agents should likewise become a focus of the community, as they introduce a distinct set of data management challenges.

\headerl{Systems and infrastructure} MLE agents repeatedly generate, execute, and evaluate a large number of candidate pipelines, making pipeline search itself a dynamic workload. This shifts the optimization scope from individual pipeline executions to the entire search process, creating opportunities for shared execution, reuse of intermediates~\cite{PhaniR021, Phani025}, cross-pipeline scheduling, and resource management across CPUs, GPUs, and distributed backends~\cite{phani2026stratum,kaoudi2026rethinkingqueryoptimizationmultiagent}. Existing work on efficient ML systems~\cite{BoehmADGIKLPR20, palkar2017weld, SujeethLBRCWAOO11} provides a natural foundation, but must be extended to workloads in which pipeline structure, code, data access patterns, resource requirements continuously evolve. Further challenges include cost- and energy-aware execution~\cite{LeisK21}, adaptive placement and parallelization~\cite{ZhengLZZCHWXZXG22}, multi-tenant resource sharing across agents~\cite{HuangB013, CarreiraFTZK19}, and tighter agent-system co-design. Finally, the LLM inference layer itself presents opportunities, such as tailored inference configuration, model selection, and reuse of key-value caches across agents~\cite{QinLHCRZ0ZX25}.

\looseness=-1
\headerl{Data discovery and preparation} Current MLE agents are largely evaluated on Kaggle-like tasks with a single pre-built input table, and therefore lack the ability to discover and prepare useful data from heterogeneous sources~\cite{yuan2026walkrunimportancedata} with complex schemas~\cite{chen2024beaver}. Real-world agents must instead operate over data warehouses and lakes, discovering relevant datasets, integrating them, and identifying useful augmentations. 
This requires integrating efficient data augmentation and discovery techniques~\cite{chepurko2020arda,huang2023kitana,turchenko2026matryoshka} into the agentic search process.
\par

\headerl{Reliability} MLE agents also face substantial reliability challenges: they often generate non-runnable or semantically incorrect pipelines~\cite{ovcharenko2026sempipes} or fail to filter and clean their input data appropriately. Addressing this requires techniques for detecting and repairing faulty code~\cite{chun2026recovering}, preventing common errors such as train-test leakage~\cite{schelter2023screening}, and data debugging methods~\cite{karlavs2022data} to identify data issues such as mislabeled training examples.

\headerl{Fairness and compliance} By automating pipeline development from raw data and natural-language instructions, MLE agents could make ML accessible to non-technical domain experts. In sensitive or regulated domains, however, this abstraction creates a responsibility gap: users may lack visibility into choices that affect correctness, robustness, fairness, and regulatory compliance~\cite{richter2026fairmachinelearningengineering,rabanser2026towards}. Open questions include how to evaluate these properties systematically and how to design agents that allow non-technical users to understand and control consequential compliance decisions.

\header{Covered technical questions} In summary, we address the following technical questions:

\begin{itemize}[leftmargin=*,noitemsep]
\item \textit{What are machine learning engineering agents, and how do they combine LLM-based code synthesis with evolutionary optimization to generate end-to-end ML pipelines?}
\item \textit{What data management challenges arise for MLE agents in runtime systems, data discovery and augmentation, reliability, as well as fairness and compliance?}
\item \textit{How can data management research contribute to addressing these challenges and improving MLE agents?}
\end{itemize}

\section{Duration, Format and Anticipated Number of Participants}

The proposed tutorial is new and has not been offered previously.

\header{Duration and expected number of participants} We propose a 1.5-hour tutorial and expect approximately 20--40 participants. The tutorial assumes only a basic understanding of machine learning and Python.

\header{Format} The tutorial consists of a 60-minute presentation followed by a 30-minute hands-on session. The presentation will combine prepared slides with interactive question answering, while the hands-on session will allow participants to explore MLE agents on their own laptops. The tutorial is structured as follows:

\headerl{Presentation (60min)} In the first 30 minutes, we introduce the core architecture of MLE agents and explain how they combine LLM-based code generation with iterative and evolutionary search to optimize end-to-end ML pipelines. We survey representative systems and benchmarks. In the second 30 minutes, we discuss open data management research questions that arise when MLE agents move beyond simplified benchmarks to real-world data environments. 

\headerl{Hands-on tutorial (30min)} Participants will experiment with several MLE agents using prepared Google Colab notebooks and observe how they generate, execute, and iteratively improve ML pipelines. The exercises will require minimal setup; participants will only need to bring a laptop and provide their own LLM API key where required.

\printbibliography
\end{document}